\documentclass[12pt,a4paper]{article}
\usepackage{a4wide}
\usepackage{latexsym}
\usepackage{epsf}
\usepackage{amssymb}
\begin{document}
\def\be{\begin{equation}}
\def\ee{\end{equation}}
\def\bea{\begin{eqnarray}}
\def\eea{\end{eqnarray}}  

\def\ads{AdS$_5\times$S$^5$ }
\def\pd{\partial}
\def\a{\alpha}
\def\b{\beta}
\def\g{\gamma}
\def\d{\delta}
\def\m{\mu}
\def\n{\nu}
\def\t{\tau} 
\def\l{\lambda}
\def\L{\Lambda}
\def\s{\sigma}
\def\e{\epsilon}
\def\scri{\mathcal{J}}
\def\cM{\mathcal{M}}
\def\tcM{\tilde{\mathcal{M}}}
\def\RR{\mathbb{R}}
\def\CC{\mathbb{C}}

\hyphenation{re-pa-ra-me-tri-za-tion}
\hyphenation{trans-for-ma-tions}


\begin{flushright}
IFT-UAM/CSIC-02-47\\
hep-th/0210274\\
\end{flushright}

\vspace{1cm}

\begin{center}

{\bf\Large The Static Gauge Potential with a Cutoff}

\vspace{.5cm}

{\bf Enrique \'Alvarez}
\footnote{E-mail: {\tt enrique.alvarez@uam.es}}
{\bf and Juan Jos\'e Manjar\'{\i}n}
\footnote{E-mail: {\tt juanjose.manjarin@uam.es}} \\
\vspace{.3cm}

\vskip 0.4cm
 {\it   
Instituto de F\'{\i}sica Te\'orica, C-XVI,
  Universidad Aut\'onoma de Madrid \\
  E-28049-Madrid, Spain}\footnote{Unidad de Investigaci\'on Asociada
  al Centro de F\'{\i}sica Miguel Catal\'an (C.S.I.C.)}

\vskip 0.2cm

and

\vskip 0.4cm
{\it Departamento de F\'{\i}sica Te\'orica, C-XI,
  Universidad Aut\'onoma de Madrid \\
  E-28049-Madrid, Spain }

\vskip 0.2cm

\vskip 1cm


{\bf Abstract}
\end{center}


\begin{quote}
The static potential, corresponding to the interaction of two heavy sources is computed for $\mathcal{N}=4$ Super Yang Mills in the strong 't Hooft coupling regime by using the AdS/CFT conjecture and performing a computation of a rectangular Wilson loop at a finite distance of the boundary.
 
\end{quote}


\newpage

\setcounter{page}{1}
\setcounter{footnote}{1}

\section{Introduction}

Since the conjecture of the equivalence between  of $d=4$  ${\mathcal{N}}=4$ super Yang-Mills and weakly  IIB string theory  on AdS$_5\times$S$^5$ with the radius of the sphere  given by $(g_{YM}^2N)^{1/4}$ in string units \cite{maldacena} and the precise recipe on how to compute correlation functions on the gauge theory using the string theory description \cite{witten, gu.kl.po}, there has been a large amount of research in order to test and to explore the consequences of this duality, (cf. \cite{aharony1})

One of the interesting computations made in this context is that of Wilson loop operators in the gauge theory \cite{wilson}. The proposal is

\be\label{su.ap}
<W(\mathcal{C})>=e^{-S},
\ee

\noindent where $S$ is the string action evaluated on-shell which, in the large $gN$ limit, becomes the area of the worldsheet, which corresponds to the interpretation of the expectation value of the Wilson loop as an statistical average over random surfaces \cite{orlando}. The potential between heavy sources in the field theory is determined by the rectangular loop defined by two infinitely heavy sources located at a distance $L$, and interacting for a very long interval of time, $T$, owing  to the fact that in the $T\rightarrow\infty$ limit, the expectation value of the Wilson loop operator takes the form

\be
<W(\mathcal{C})>=A(L)e^{-TE(L)},
\ee

\noindent We then will parametrize the interaction of the sources in terms of the distance between them via the relation

\be
E(L)=\lim_{T\rightarrow\infty}\frac{S}{T}.
\ee

In general, one should take the whole string partition function of type IIB on \ads but the semiclassical level of approximation is believed to make sense in this particular example unless the integration over \ads has infrared divergences, in which case one should include stringy corrections to (\ref{su.ap}) \cite{witten}.

The apearance of these divergences can be associated to anomalies of the conformal field theory. In this context, it can be seen that the IR regularization of the theory in the bulk corresponds to an UV regularization on the gauge theory \cite{sus.wi}, (the so called IR-UV connection). To be specific, we can consider the case of the dilaton, whose expectation value is set by its boundary conditions at infinity so that changing the gauge theory coupling constant corresponds to changing the boundary value of the dilaton. In this sense, different regions of the AdS located at different points of the radial coordinate correspond to different energy scales in the field theory.

Is precisely this connection the main question of this paper, which we will examine in the simplest context of type IIB and its relation with $\mathcal{N}=4$ SYM, a nonrealistic theory. This analysis can presumably be pushed forward to encompass the non conformal case \cite{ec1,kogan,al.man}. However, the analysis done here completes the one of \cite{al.man} and allows the study with, some more detail, of the decoupling limit of the gauge theory, that is, the limit in which there is no interaction with the gravity of the bulk (\cite{Gubser}).

In order to study the dependence of the gauge theory with the regulator of the theory in the bulk, we will compute the potential between two heavy sources. The static limit of this configuration is obtained by taking the masses of the particles to infinity. 

We will begin with a stack of $N$ D3-branes and Higgs down the system to  $N-1$ D3-branes located at the origin and one D3-brane located at an arbitrary point in the radial direction. Is in this separated brane where the sources are placed, which are nothing but the end points of an open string attached to the brane. From this point of view, the infinite mass of the particles corresponds to the limit in which the brane is infinitely far away.

This configuration obviously leads to a divergence in the potential which can be understood as coming from the fact that the string goes from the system of branes to the brane at infinity and must be renormalized in such a way that we are left with a single string with its two ends attached to the brane at infinity and the embedding resembles topologically half a cylinder, a $U$-like configuration. 

In section \ref{cut.pot} we will compute the potential and its dependence with the distance between the heavy sources and the regulator. In section \ref{comm} we will explore the meaning of this dependence and we will end up with some comments concerning the decoupling limit.

\section{Cutoff dependence of the static potential}\label{cut.pot}
In order to compute the static gauge potential we will use the form of the AdS$_5\times$S$^5$ which makes clear its conformal relation to flat space, namely

\be
\label{ads}
ds^2=a(y)\left( dx^2_4+dy^2\right),
\ee

\noindent where $a(y)=\frac{\a'R^2}{y^2}$ and $y$ has units of length. The metric induced on the worldsheet of the string is now

\be
ds^2_{ind}=\left[ a(y)\left( (x')^2+(y')^2\right)\right]d\s^2+a(y)(\dot t)^2d\t^2,
\ee

\noindent where $(\s,\t)$ parametrize the 2d string worldsheet and where the prime stands for a derivative with respect to sigma and the dot with respect to tau. Now we can write the Polyakov action as

\be
\label{pol.ac}
S_p=\frac{R^2}{4\pi}\int_M \frac{(x')^2+(y')^2+T^2}{y^2}d^2\Sigma,
\ee

\noindent where we have parametrized $t=\t T$. The conserved quantities asociated with this action principle are the canonical momentum conjugate to $x$ and the energy-momentum tensor,

\be
p=\frac{R^2}{2\pi}\frac{x'}{y^2},
\ee
\be
j=\frac{R^2}{4\pi}\frac{(x')^2+(y')^2-T^2}{y^2}.
\ee

In this context we still need to choose the conformal gauge in order to turn (\ref{pol.ac}) into the Dirichlet functional whose on-shell solutions will give us the minimal area surfaces. The Virasoro constraints read 

\be
\label{vir.con}
j=0,\qquad p^2=\frac{T^2}{y_0^4} ,
\ee

\noindent where $y_0$ is the the tip of the $U$-like configuration of the string, defined through the condition $y'=0$. 

Once we have chosen this gauge, we are  left with no more freedom which in turn implies that the range of the parameter $\s$ remains to be determined. It is worth to mention that when working with the Nambu-Goto action, one can choose the static gauge, which fixes $\s=x$ and $\t=t$. 

With conditions (\ref{vir.con}) we have, then, three integrals to solve, which determine the range of $\s$, the length and the energy of the interaction between the heavy sources. These read 

\be
\mathfrak{R}(\s)=\int d\s=\int \frac{d\s}{dy}dy=\frac{y_0}{T}\int_1^{\Lambda/y_0}\frac{d\xi}{\sqrt{(1-\xi^2)(1+\xi^2)}},
\ee
\be
L=\int dy =\int \frac{dx}{d\s}\frac{d\s}{dy}dy=y_0^2\int_1^{\Lambda/y_0}\frac{\xi^2d\xi}{\sqrt{(1-\xi^2)(1+\xi^2)}},
\ee
\be
S_p=\frac{R^2T}{2\pi y_0}\int_1^{\Lambda/y_0}\frac{d\xi}{\xi^2\sqrt{(1-\xi^2)(1+\xi^2)}},
\ee

\noindent where $\Lambda$ is the point in the $y$ direction where we locate the D3-brane to compute the potential and is believed to act as a cutoff for the conformal theory as we will explore later.

These integrals can be easily written as a combination of elliptic functions

\be
\mathfrak{R}(\s)=-\frac{y_0}{\sqrt{2}T}F\left(\cos^{-1}\left(\Lambda/y_0\right),1/\sqrt{2}\right),
\ee
\be
\label{int.l}
L=-\sqrt{2}y_0^2\left[ F\left(\cos^{-1}\left(\Lambda/y_0\right),1/\sqrt{2}\right)-2E\left(\cos^{-1}\left(\Lambda/y_0\right),1/\sqrt{2}\right)\right],
\ee
\be
\label{int.pot}
S_p=\frac{R^2T}{2\pi y_0}\left\{\frac{1}{\sqrt{2}}\left[ F\left(\cos^{-1}\left(\Lambda/y_0\right),1/\sqrt{2}\right)-2E\left(\cos^{-1}\left(\Lambda/y_0\right),1/\sqrt{2}\right)\right]+\frac{y_0}{\Lambda}\sqrt{1-\frac{\Lambda^4}{y_0^4}}\right\}.
\ee

The final goal of this section is to find the explicit dependence of the potential with the distance between the heavy sources and the cutoff $\Lambda$. We will use the aproximation of \cite{al.man}, which allows us to write 

\be
\label{ran.sig}
\mathfrak{R}(\s)=\frac{y_0}{T\sqrt{3}}\sqrt{\frac{\Lambda^2}{y_0^2}-5\frac{\Lambda}{y_0}+4},
\ee

\noindent and

\be
V=\frac{R^2}{2\pi y_0}\left(\sqrt{\frac{y_0^2}{\Lambda^2}-\frac{\Lambda^2}{y_0^2}}-\sqrt{\frac{\Lambda}{y_0}-\frac{\Lambda^2}{y_0^2}}\right).
\ee

This last expresion needs to be renormalized because it diverges linearly in the $\Lambda\rightarrow 0$ limit, as seen also from (\ref{int.pot}), that is, as we move towards the boundary of AdS$_5$. This divergence can be undestood as an infinite self-energy of the string attached to the branes and the corresponding renormalization as a mass renormalization. 

The renormalized potential will be obtained from

\be
V=\frac{R^2}{2\pi y_0}\int_1^{\Lambda/y_0}\left(\frac{1}{\xi^2\sqrt{(1-\xi^2)(1+\xi^2)}}+\frac{1}{\xi^2}\right)d\xi,
\ee

\noindent which we will write, in our approximation, as

\be
\label{pot.ren}
V=\frac{R^2}{2\pi y_0}\left(\sqrt{\frac{y_0^2}{\Lambda^2}-\frac{\Lambda^2}{y_0^2}}-\sqrt{\frac{\Lambda}{y_0}-\frac{\Lambda^2}{y_0^2}}+1-\frac{y_0}{\Lambda}\right).
\ee

\begin{figure}[h]
\begin{center}
\leavevmode
\epsfxsize=5cm
\epsffile{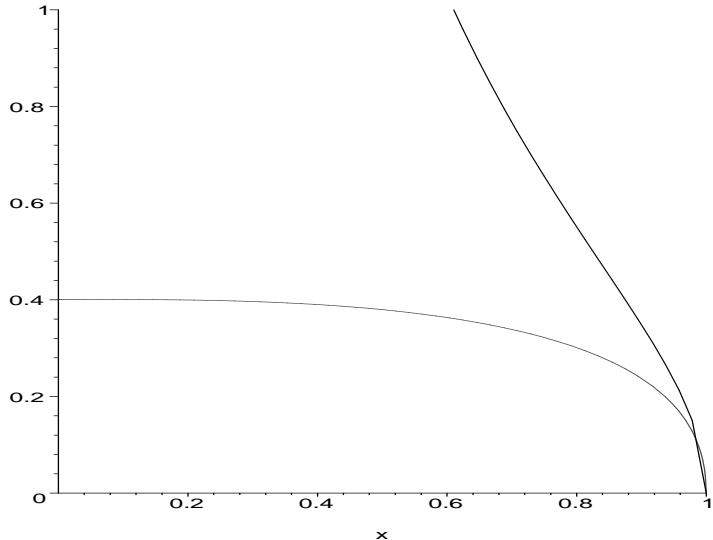}
\caption{\it Comparison between the renormalized (lightest line) and the non-renormalized (darkest line) potentials. In this plot $x=\Lambda/y_0$ and $y=2\pi y_0V/R^2$}
\label{f2}
\end{center}
\end{figure}

\begin{figure}[h]
\begin{center}
\leavevmode
\epsfxsize=5cm
\epsffile{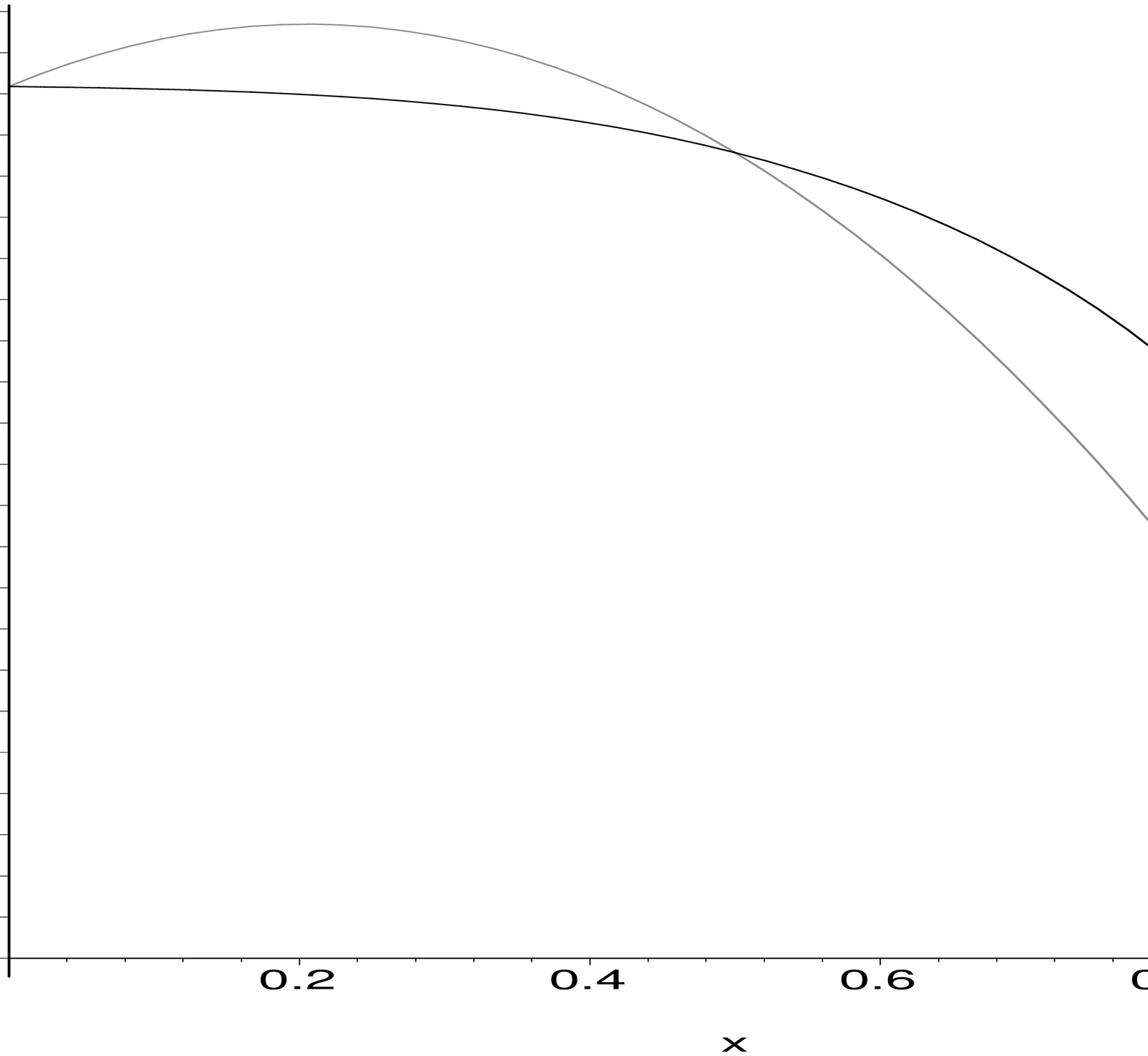}
\caption{\it Comparison between the exact (darkest line) and the second order interpolation (lightest line), where $x=\Lambda/y_0$ and $y=L/\sqrt{2}y_0^2$}
\label{f1}
\end{center}
\end{figure}

In the case of (\ref{int.l}), we cannot use the same approximation due to the absence of the divergent term, which causes the approximation to fail in the $\Lambda/y_0\rightarrow 0$ limit. In this case we can make a polynomial interpolation with fixed end points at $\Lambda/y_0=0$ and $\Lambda/y_0=1$ which, to second order this reads 

\be
L=\sqrt{2}y_0\left(-1.44\frac{\Lambda^2}{y_0^2}+0.59\frac{\Lambda}{y_0}+0.85\right),
\ee

\noindent which will introduce a slight deviation from the exact curve small enough to  allow us to make accurate predictions, see fig. (\ref{f1}). This lets us to solve $y_0$ as a function of $\Lambda$ and $L$ as

\be
\label{y.0}
y_0=-0.35\Lambda+1.35\sqrt{\Lambda^2+0.46L}.
\ee

Substitution of (\ref{y.0}) into (\ref{ran.sig}) and (\ref{pot.ren}) will then produce the desired expresions for the potential and the range of $\s$ in terms of the physical parameters $L$ and $\Lambda$.

\section{Concluding remarks on the meaning of the cutoff}\label{comm}
In the previous section we have used the usual construction to compute the potential between heavy sources \cite{wilson}. Starting with a stack of $N+1$ D3-branes, which enjoy a low energy description in terms of the $U(N+1)$ theory living on them,  we Higgs it down to a system of $N$ D3-branes and a D3-brane located at a position $\Lambda$ on the radial direction $y$ of (\ref{ads}). This breaks the gauge group down to $U(N)\times U(1)$ and can be understood as giving an expectation value to a scalar field, $\phi$ of the theory, in such a way that at very short distances $<<1/\phi$ the gauge group remains unbroken. All this implies that the infinite mass limit, $\phi\rightarrow\infty$, is necessary to compute the Wilson loop of the $U(N)$ gauge theory.

\begin{figure}[h]
\begin{center}
\leavevmode
\epsfxsize=8cm
\epsffile{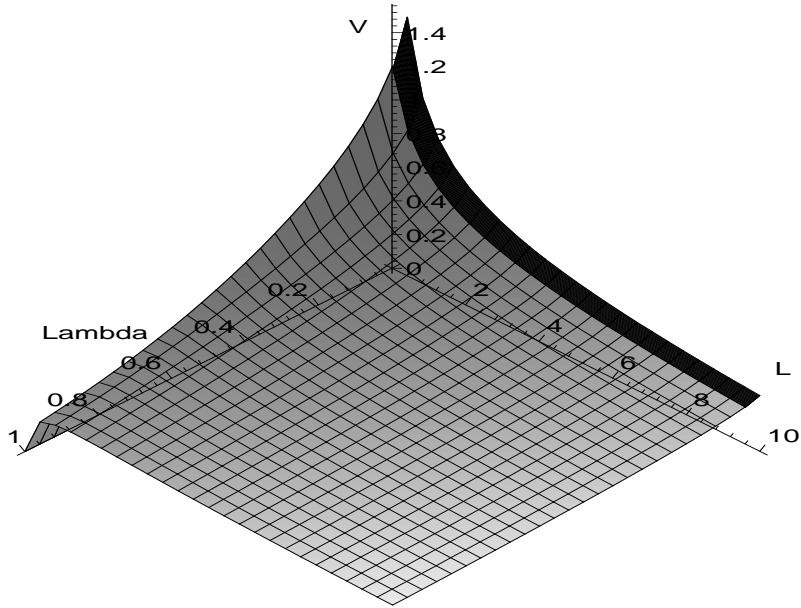}
\caption{\it Three dimensional representation of the potential as a function of the distantce between the heavy W-bosons on the brane and the position of the brane on the radial direction}
\label{f3}
\end{center}
\end{figure}

In fig. (\ref{f3}) we have represented the renormalized potential (\ref{pot.ren}) in terms of $L$ and $\Lambda$ and we can see that is precisely in the boundary where the conformal theory lives, that is, only when $\Lambda=0$ we find a $1/L$ dependence in the potential.

At this point there is one important question that is, what information is encoding this potential about the field theory on the brane?. The standard belief is that this setup is equivalent to a CFT with a cutoff, coupled to gravity.We do not find any divergence, which is only natural in a CFT with vanishing beta function. But it is not easy to disentangle the effects of the cutoff from the effects of gravity itself (which should change the coefficient of the $1/L$ term and, in addition, add a new term in $1/L^3$). Another interesting feature of this potential is that far enough from the boundary describes a plateau where, for every value of $\Lambda$ and $L$, it is always zero.

It is difficult though to gauge the exact physical meaning of this result, owing to the fact that  finite cutoff corrections can disguise themselves in unsuspected ways, specially when some approximation is made. Consider, for the sake of the argument, the (finite) integral

\be
V(L)=\int_0^{\infty} dk e^{-k L}=\frac{1}{L}.
\ee

If now a cutoff is introduced,

\be
V(\Lambda,L)=\int_0^{\Lambda} dk e^{-k L}=\frac{1}{L}(1-e^{-\Lambda L}).
\ee

When $\Lambda L\rightarrow\infty$ the limit of $V(\Lambda,L)$ towards $V(L)$ is smooth. But if, for example, an expansion is made for $\Lambda L<< 1$, it would result in

\be
V(\Lambda,L)\sim \Lambda(1-\frac{\Lambda L}{2}),
\ee

\noindent which has a zero at $\Lambda L=2$, outside the region of validity of the Taylor expansion.

\begin{figure}[h]
\begin{center}
\leavevmode
\epsfxsize=8cm
\epsffile{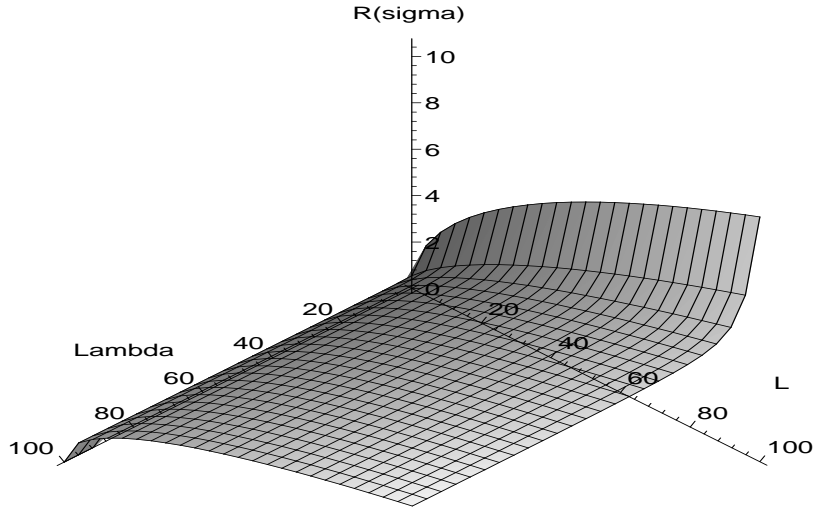}
\caption{\it Three dimensional representation of sigma as a function of the distantce between the heavy W-bosons on the brane and the position of the brane on the radial direction}
\label{f4}
\end{center}
\end{figure}

The range of  $\sigma$, that now is dinamically determined, has been also plotted for completeness.

Work is in progress to extend this computations to different backgrounds purpoted to represent non-conformal situations.

\section*{Acknowledgments}
We would like to thank M. Garc\'\i a-P\'erez for useful conversations. This work has been partially supported by the European Commission (HPRN-CT-200-00148) and CICYT (Spain).      


\end{document}